\documentclass[twocolumn,appendixfloats,twocolappendix]{aastex631}

\usepackage{natbib}
\usepackage{amsmath,amssymb}
\usepackage{xspace}
\usepackage{rotating}
\usepackage{multirow}
\usepackage{enumitem}
\usepackage{pifont}
\usepackage{bm} 
\newcommand{\vect}[1]{\boldsymbol{#1}}
\usepackage{dcolumn}
\usepackage{macros}
\usepackage{xcolor,hyperref}
\usepackage{fdsymbol}
\usepackage{graphicx} 
\definecolor{mpl_blue}{HTML}{1F77B4}
\definecolor{mpl_orange}{HTML}{FF7F0E}
\definecolor{mpl_green}{HTML}{2CA02C}
\definecolor{mpl_red}{HTML}{D62728}

\usepackage[all]{hypcap}
\usepackage[caption=false]{subfig}
\newcommand{\m}[1]{\mathbf{#1}}

\def\be{\begin{equation}}
\def\ee{\end{equation}}
\newcommand{\bb}{\begin{bmatrix}}
\newcommand{\eb}{\end{bmatrix}}
\def\bea{\begin{eqnarray}}
\def\eea{\end{eqnarray}}

\def\R{\color{red}}

\defcitealias{15yrGWB}{NG15}
\defcitealias{NG}{NG12.5}

\begin{document}
\title{The NANOGrav 15-year data set: Search for Transverse Polarization Modes in the Gravitational-Wave Background }

\shorttitle{NANOGrav 15-year Gravitational-Wave Background}
\shortauthors{The NANOGrav Collaboration}

\author[0000-0001-5134-3925]{Gabriella Agazie}
\affiliation{Center for Gravitation, Cosmology and Astrophysics, Department of Physics, University of Wisconsin-Milwaukee,\\ P.O. Box 413, Milwaukee, WI 53201, USA}
\author[0000-0002-8935-9882]{Akash Anumarlapudi}
\affiliation{Center for Gravitation, Cosmology and Astrophysics, Department of Physics, University of Wisconsin-Milwaukee,\\ P.O. Box 413, Milwaukee, WI 53201, USA}
\author[0000-0003-0638-3340]{Anne M. Archibald}
\affiliation{Newcastle University, NE1 7RU, UK}
\author{Zaven Arzoumanian}
\affiliation{X-Ray Astrophysics Laboratory, NASA Goddard Space Flight Center, Code 662, Greenbelt, MD 20771, USA}
\author{Jeremy Baier}
\affiliation{Department of Physics, Oregon State University, Corvallis, OR 97331, USA}
\author[0000-0003-2745-753X]{Paul T. Baker}
\affiliation{Department of Physics and Astronomy, Widener University, One University Place, Chester, PA 19013, USA}
\author[0000-0003-0909-5563]{Bence B\'{e}csy}
\affiliation{Department of Physics, Oregon State University, Corvallis, OR 97331, USA}
\author[0000-0002-2183-1087]{Laura Blecha}
\affiliation{Physics Department, University of Florida, Gainesville, FL 32611, USA}
\author[0000-0001-6341-7178]{Adam Brazier}
\affiliation{Cornell Center for Astrophysics and Planetary Science and Department of Astronomy, Cornell University, Ithaca, NY 14853, USA}
\affiliation{Cornell Center for Advanced Computing, Cornell University, Ithaca, NY 14853, USA}
\author[0000-0003-3053-6538]{Paul R. Brook}
\affiliation{Institute for Gravitational Wave Astronomy and School of Physics and Astronomy, University of Birmingham, Edgbaston, Birmingham B15 2TT, UK}
\author[0000-0003-4052-7838]{Sarah Burke-Spolaor}
\affiliation{Department of Physics and Astronomy, West Virginia University, P.O. Box 6315, Morgantown, WV 26506, USA}
\affiliation{Center for Gravitational Waves and Cosmology, West Virginia University, Chestnut Ridge Research Building, Morgantown, WV 26505, USA}
\author{Rand Burnette}
\affiliation{Department of Physics, Oregon State University, Corvallis, OR 97331, USA}
\author{Robin Case}
\affiliation{Department of Physics, Oregon State University, Corvallis, OR 97331, USA}
\author[0000-0002-5557-4007]{J. Andrew Casey-Clyde}
\affiliation{Department of Physics, University of Connecticut, 196 Auditorium Road, U-3046, Storrs, CT 06269-3046, USA}
\author[0000-0003-3579-2522]{Maria Charisi}
\affiliation{Department of Physics and Astronomy, Vanderbilt University, 2301 Vanderbilt Place, Nashville, TN 37235, USA}
\author[0000-0002-2878-1502]{Shami Chatterjee}
\affiliation{Cornell Center for Astrophysics and Planetary Science and Department of Astronomy, Cornell University, Ithaca, NY 14853, USA}
\author[0000-0001-7587-5483]{Tyler Cohen}
\affiliation{Department of Physics, New Mexico Institute of Mining and Technology, 801 Leroy Place, Socorro, NM 87801, USA}
\author[0000-0002-4049-1882]{James M. Cordes}
\affiliation{Cornell Center for Astrophysics and Planetary Science and Department of Astronomy, Cornell University, Ithaca, NY 14853, USA}
\author[0000-0002-7435-0869]{Neil J. Cornish}
\affiliation{Department of Physics, Montana State University, Bozeman, MT 59717, USA}
\author[0000-0002-2578-0360]{Fronefield Crawford}
\affiliation{Department of Physics and Astronomy, Franklin \& Marshall College, P.O. Box 3003, Lancaster, PA 17604, USA}
\author[0000-0002-6039-692X]{H. Thankful Cromartie}
\altaffiliation{NASA Hubble Fellowship: Einstein Postdoctoral Fellow}
\affiliation{Cornell Center for Astrophysics and Planetary Science and Department of Astronomy, Cornell University, Ithaca, NY 14853, USA}
\author[0000-0002-1529-5169]{Kathryn Crowter}
\affiliation{Department of Physics and Astronomy, University of British Columbia, 6224 Agricultural Road, Vancouver, BC V6T 1Z1, Canada}
\author[0000-0002-2185-1790]{Megan E. DeCesar}
\affiliation{George Mason University, resident at the Naval Research Laboratory, Washington, DC 20375, USA}
\author[0009-0009-3479-9897]{Dallas DeGan}
\affiliation{Department of Physics, Oregon State University, Corvallis, OR 97331, USA}
\author[0000-0002-6664-965X]{Paul B. Demorest}
\affiliation{National Radio Astronomy Observatory, 1003 Lopezville Rd., Socorro, NM 87801, USA}
\author[0000-0001-8885-6388]{Timothy Dolch}
\affiliation{Department of Physics, Hillsdale College, 33 E. College Street, Hillsdale, MI 49242, USA}
\affiliation{Eureka Scientific, 2452 Delmer Street, Suite 100, Oakland, CA 94602-3017, USA}
\author{Brendan Drachler}
\affiliation{School of Physics and Astronomy, Rochester Institute of Technology, Rochester, NY 14623, USA}
\affiliation{Laboratory for Multiwavelength Astrophysics, Rochester Institute of Technology, Rochester, NY 14623, USA}
\author[0000-0001-7828-7708]{Elizabeth C. Ferrara}
\affiliation{Department of Astronomy, University of Maryland, College Park, MD 20742}
\affiliation{Center for Research and Exploration in Space Science and Technology, NASA/GSFC, Greenbelt, MD 20771}
\affiliation{NASA Goddard Space Flight Center, Greenbelt, MD 20771, USA}
\author[0000-0001-5645-5336]{William Fiore}
\affiliation{Department of Physics and Astronomy, West Virginia University, P.O. Box 6315, Morgantown, WV 26506, USA}
\affiliation{Center for Gravitational Waves and Cosmology, West Virginia University, Chestnut Ridge Research Building, Morgantown, WV 26505, USA}
\author[0000-0001-8384-5049]{Emmanuel Fonseca}
\affiliation{Department of Physics and Astronomy, West Virginia University, P.O. Box 6315, Morgantown, WV 26506, USA}
\affiliation{Center for Gravitational Waves and Cosmology, West Virginia University, Chestnut Ridge Research Building, Morgantown, WV 26505, USA}
\author[0000-0001-7624-4616]{Gabriel E. Freedman}
\affiliation{Center for Gravitation, Cosmology and Astrophysics, Department of Physics, University of Wisconsin-Milwaukee,\\ P.O. Box 413, Milwaukee, WI 53201, USA}
\author[0000-0001-6166-9646]{Nate Garver-Daniels}
\affiliation{Department of Physics and Astronomy, West Virginia University, P.O. Box 6315, Morgantown, WV 26506, USA}
\affiliation{Center for Gravitational Waves and Cosmology, West Virginia University, Chestnut Ridge Research Building, Morgantown, WV 26505, USA}
\author[0000-0001-8158-683X]{Peter A. Gentile}
\affiliation{Department of Physics and Astronomy, West Virginia University, P.O. Box 6315, Morgantown, WV 26506, USA}
\affiliation{Center for Gravitational Waves and Cosmology, West Virginia University, Chestnut Ridge Research Building, Morgantown, WV 26505, USA}
\author[0000-0003-4090-9780]{Joseph Glaser}
\affiliation{Department of Physics and Astronomy, West Virginia University, P.O. Box 6315, Morgantown, WV 26506, USA}
\affiliation{Center for Gravitational Waves and Cosmology, West Virginia University, Chestnut Ridge Research Building, Morgantown, WV 26505, USA}
\author[0000-0003-1884-348X]{Deborah C. Good}
\affiliation{Department of Physics, University of Connecticut, 196 Auditorium Road, U-3046, Storrs, CT 06269-3046, USA}
\affiliation{Center for Computational Astrophysics, Flatiron Institute, 162 5th Avenue, New York, NY 10010, USA}
\author[0000-0002-1146-0198]{Kayhan G\"{u}ltekin}
\affiliation{Department of Astronomy and Astrophysics, University of Michigan, Ann Arbor, MI 48109, USA}
\author[0000-0003-2742-3321]{Jeffrey S. Hazboun}
\affiliation{Department of Physics, Oregon State University, Corvallis, OR 97331, USA}
\author[0000-0003-1082-2342]{Ross J. Jennings}
\altaffiliation{NANOGrav Physics Frontiers Center Postdoctoral Fellow}
\affiliation{Department of Physics and Astronomy, West Virginia University, P.O. Box 6315, Morgantown, WV 26506, USA}
\affiliation{Center for Gravitational Waves and Cosmology, West Virginia University, Chestnut Ridge Research Building, Morgantown, WV 26505, USA}
\author[0000-0002-7445-8423]{Aaron D. Johnson}
\affiliation{Center for Gravitation, Cosmology and Astrophysics, Department of Physics, University of Wisconsin-Milwaukee,\\ P.O. Box 413, Milwaukee, WI 53201, USA}
\affiliation{Division of Physics, Mathematics, and Astronomy, California Institute of Technology, Pasadena, CA 91125, USA}
\author[0000-0001-6607-3710]{Megan L. Jones}
\affiliation{Center for Gravitation, Cosmology and Astrophysics, Department of Physics, University of Wisconsin-Milwaukee,\\ P.O. Box 413, Milwaukee, WI 53201, USA}
\author[0000-0002-3654-980X]{Andrew R. Kaiser}
\affiliation{Department of Physics and Astronomy, West Virginia University, P.O. Box 6315, Morgantown, WV 26506, USA}
\affiliation{Center for Gravitational Waves and Cosmology, West Virginia University, Chestnut Ridge Research Building, Morgantown, WV 26505, USA}
\author[0000-0001-6295-2881]{David L. Kaplan}
\affiliation{Center for Gravitation, Cosmology and Astrophysics, Department of Physics, University of Wisconsin-Milwaukee,\\ P.O. Box 413, Milwaukee, WI 53201, USA}
\author[0000-0002-6625-6450]{Luke Zoltan Kelley}
\affiliation{Department of Astronomy, University of California, Berkeley, 501 Campbell Hall \#3411, Berkeley, CA 94720, USA}
\author[0000-0002-0893-4073]{Matthew Kerr}
\affiliation{Space Science Division, Naval Research Laboratory, Washington, DC 20375-5352, USA}
\author[0000-0003-0123-7600]{Joey S. Key}
\affiliation{University of Washington Bothell, 18115 Campus Way NE, Bothell, WA 98011, USA}
\author[0000-0002-9197-7604]{Nima Laal}
\affiliation{Department of Physics, Oregon State University, Corvallis, OR 97331, USA}
\author[0000-0003-0721-651X]{Michael T. Lam}
\affiliation{School of Physics and Astronomy, Rochester Institute of Technology, Rochester, NY 14623, USA}
\affiliation{Laboratory for Multiwavelength Astrophysics, Rochester Institute of Technology, Rochester, NY 14623, USA}
\author[0000-0003-1096-4156]{William G. Lamb}
\affiliation{Department of Physics and Astronomy, Vanderbilt University, 2301 Vanderbilt Place, Nashville, TN 37235, USA}
\author{T. Joseph W. Lazio}
\affiliation{Jet Propulsion Laboratory, California Institute of Technology, 4800 Oak Grove Drive, Pasadena, CA 91109, USA}
\author[0000-0003-0771-6581]{Natalia Lewandowska}
\affiliation{Department of Physics, State University of New York at Oswego, Oswego, NY, 13126, USA}
\author[0000-0001-5766-4287]{Tingting Liu}
\affiliation{Department of Physics and Astronomy, West Virginia University, P.O. Box 6315, Morgantown, WV 26506, USA}
\affiliation{Center for Gravitational Waves and Cosmology, West Virginia University, Chestnut Ridge Research Building, Morgantown, WV 26505, USA}
\author[0000-0003-1301-966X]{Duncan R. Lorimer}
\affiliation{Department of Physics and Astronomy, West Virginia University, P.O. Box 6315, Morgantown, WV 26506, USA}
\affiliation{Center for Gravitational Waves and Cosmology, West Virginia University, Chestnut Ridge Research Building, Morgantown, WV 26505, USA}
\author[0000-0001-5373-5914]{Jing Luo}
\altaffiliation{Deceased}
\affiliation{Department of Astronomy \& Astrophysics, University of Toronto, 50 Saint George Street, Toronto, ON M5S 3H4, Canada}
\author[0000-0001-5229-7430]{Ryan S. Lynch}
\affiliation{Green Bank Observatory, P.O. Box 2, Green Bank, WV 24944, USA}
\author[0000-0002-4430-102X]{Chung-Pei Ma}
\affiliation{Department of Astronomy, University of California, Berkeley, 501 Campbell Hall \#3411, Berkeley, CA 94720, USA}
\affiliation{Department of Physics, University of California, Berkeley, CA 94720, USA}
\author[0000-0003-2285-0404]{Dustin R. Madison}
\affiliation{Department of Physics, University of the Pacific, 3601 Pacific Avenue, Stockton, CA 95211, USA}
\author[0000-0001-5481-7559]{Alexander McEwen}
\affiliation{Center for Gravitation, Cosmology and Astrophysics, Department of Physics, University of Wisconsin-Milwaukee,\\ P.O. Box 413, Milwaukee, WI 53201, USA}
\author[0000-0002-2885-8485]{James W. McKee}
\affiliation{E.A. Milne Centre for Astrophysics, University of Hull, Cottingham Road, Kingston-upon-Hull, HU6 7RX, UK}
\affiliation{Centre of Excellence for Data Science, Artificial Intelligence and Modelling (DAIM), University of Hull, Cottingham Road, Kingston-upon-Hull, HU6 7RX, UK}
\author[0000-0001-7697-7422]{Maura A. McLaughlin}
\affiliation{Department of Physics and Astronomy, West Virginia University, P.O. Box 6315, Morgantown, WV 26506, USA}
\affiliation{Center for Gravitational Waves and Cosmology, West Virginia University, Chestnut Ridge Research Building, Morgantown, WV 26505, USA}
\author[0000-0002-4642-1260]{Natasha McMann}
\affiliation{Department of Physics and Astronomy, Vanderbilt University, 2301 Vanderbilt Place, Nashville, TN 37235, USA}
\author[0000-0001-8845-1225]{Bradley W. Meyers}
\affiliation{Department of Physics and Astronomy, University of British Columbia, 6224 Agricultural Road, Vancouver, BC V6T 1Z1, Canada}
\affiliation{International Centre for Radio Astronomy Research, Curtin University, Bentley, WA 6102, Australia}
\author[0000-0002-4307-1322]{Chiara M. F. Mingarelli}
\affiliation{Center for Computational Astrophysics, Flatiron Institute, 162 5th Avenue, New York, NY 10010, USA}
\affiliation{Department of Physics, University of Connecticut, 196 Auditorium Road, U-3046, Storrs, CT 06269-3046, USA}
\affiliation{Department of Physics, Yale University, New Haven, CT 06520, USA}
\author[0000-0003-2898-5844]{Andrea Mitridate}
\affiliation{Deutsches Elektronen-Synchrotron DESY, Notkestr. 85, 22607 Hamburg, Germany}
\author{Priyamvada Natarajan}
\affiliation{Department of Astronomy, Yale University, 52 Hillhouse Ave., New Haven, CT 06511, USA}
\affiliation{Black Hole Initiative, Harvard University, 20 Garden Street, Cambridge, MA 02138, USA}
\author[0000-0002-3616-5160]{Cherry Ng}
\affiliation{Dunlap Institute for Astronomy and Astrophysics, University of Toronto, 50 St. George St., Toronto, ON M5S 3H4, Canada}
\author[0000-0002-6709-2566]{David J. Nice}
\affiliation{Department of Physics, Lafayette College, Easton, PA 18042, USA}
\author[0000-0002-4941-5333]{Stella Koch Ocker}
\affiliation{Cornell Center for Astrophysics and Planetary Science and Department of Astronomy, Cornell University, Ithaca, NY 14853, USA}
\author[0000-0002-2027-3714]{Ken D. Olum}
\affiliation{Institute of Cosmology, Department of Physics and Astronomy, Tufts University, Medford, MA 02155, USA}
\author[0000-0001-5465-2889]{Timothy T. Pennucci}
\affiliation{Institute of Physics and Astronomy, E\"{o}tv\"{o}s Lor\'{a}nd University, P\'{a}zm\'{a}ny P. s. 1/A, 1117 Budapest, Hungary}
\author[0000-0002-8509-5947]{Benetge B. P. Perera}
\affiliation{Arecibo Observatory, HC3 Box 53995, Arecibo, PR 00612, USA}
\author[0000-0002-8826-1285]{Nihan S. Pol}
\affiliation{Department of Physics and Astronomy, Vanderbilt University, 2301 Vanderbilt Place, Nashville, TN 37235, USA}
\author[0000-0002-2074-4360]{Henri A. Radovan}
\affiliation{Department of Physics, University of Puerto Rico, Mayag\"{u}ez, PR 00681, USA}
\author[0000-0001-5799-9714]{Scott M. Ransom}
\affiliation{National Radio Astronomy Observatory, 520 Edgemont Road, Charlottesville, VA 22903, USA}
\author[0000-0002-5297-5278]{Paul S. Ray}
\affiliation{Space Science Division, Naval Research Laboratory, Washington, DC 20375-5352, USA}
\author[0000-0003-4915-3246]{Joseph D. Romano}
\affiliation{Department of Physics, Texas Tech University, Box 41051, Lubbock, TX 79409, USA}
\author[0000-0001-7832-9066]{Alexander Saffer}
\affiliation{Department of Physics and Astronomy, West Virginia University, P.O. Box 6315, Morgantown, WV 26506, USA}
\affiliation{Center for Gravitational Waves and Cosmology, West Virginia University, Chestnut Ridge Research Building, Morgantown, WV 26505, USA}
\author[0009-0006-5476-3603]{Shashwat C. Sardesai}
\affiliation{Center for Gravitation, Cosmology and Astrophysics, Department of Physics, University of Wisconsin-Milwaukee,\\ P.O. Box 413, Milwaukee, WI 53201, USA}
\author[0000-0003-4391-936X]{Ann Schmiedekamp}
\affiliation{Department of Physics, Penn State Abington, Abington, PA 19001, USA}
\author[0000-0002-1283-2184]{Carl Schmiedekamp}
\affiliation{Department of Physics, Penn State Abington, Abington, PA 19001, USA}
\author[0000-0003-2807-6472]{Kai Schmitz}
\affiliation{Institute for Theoretical Physics, University of M\"{u}nster, 48149 M\"{u}nster, Germany}
\author[0000-0002-7283-1124]{Brent J. Shapiro-Albert}
\affiliation{Department of Physics and Astronomy, West Virginia University, P.O. Box 6315, Morgantown, WV 26506, USA}
\affiliation{Center for Gravitational Waves and Cosmology, West Virginia University, Chestnut Ridge Research Building, Morgantown, WV 26505, USA}
\affiliation{Giant Army, 915A 17th Ave, Seattle WA 98122}
\author[0000-0002-7778-2990]{Xavier Siemens}
\affiliation{Department of Physics, Oregon State University, Corvallis, OR 97331, USA}
\affiliation{Center for Gravitation, Cosmology and Astrophysics, Department of Physics, University of Wisconsin-Milwaukee,\\ P.O. Box 413, Milwaukee, WI 53201, USA}
\author[0000-0003-1407-6607]{Joseph Simon}
\altaffiliation{NSF Astronomy and Astrophysics Postdoctoral Fellow}
\affiliation{Department of Astrophysical and Planetary Sciences, University of Colorado, Boulder, CO 80309, USA}
\author[0000-0002-1530-9778]{Magdalena S. Siwek}
\affiliation{Center for Astrophysics, Harvard University, 60 Garden St, Cambridge, MA 02138}
\author[0000-0001-9784-8670]{Ingrid H. Stairs}
\affiliation{Department of Physics and Astronomy, University of British Columbia, 6224 Agricultural Road, Vancouver, BC V6T 1Z1, Canada}
\author[0000-0002-1797-3277]{Daniel R. Stinebring}
\affiliation{Department of Physics and Astronomy, Oberlin College, Oberlin, OH 44074, USA}
\author[0000-0002-7261-594X]{Kevin Stovall}
\affiliation{National Radio Astronomy Observatory, 1003 Lopezville Rd., Socorro, NM 87801, USA}
\author[0000-0002-7933-493X]{Jerry P. Sun}
\affiliation{Department of Physics, Oregon State University, Corvallis, OR 97331, USA}
\author[0000-0002-2820-0931]{Abhimanyu Susobhanan}
\affiliation{Center for Gravitation, Cosmology and Astrophysics, Department of Physics, University of Wisconsin-Milwaukee,\\ P.O. Box 413, Milwaukee, WI 53201, USA}
\author[0000-0002-1075-3837]{Joseph K. Swiggum}
\altaffiliation{NANOGrav Physics Frontiers Center Postdoctoral Fellow}
\affiliation{Department of Physics, Lafayette College, Easton, PA 18042, USA}
\author[0000-0001-9118-5589]{Jacob A. Taylor}
\affiliation{Department of Physics, Oregon State University, Corvallis, OR 97331, USA}
\author[0000-0003-0264-1453]{Stephen R. Taylor}
\affiliation{Department of Physics and Astronomy, Vanderbilt University, 2301 Vanderbilt Place, Nashville, TN 37235, USA}
\author[0000-0002-2451-7288]{Jacob E. Turner}
\affiliation{Department of Physics and Astronomy, West Virginia University, P.O. Box 6315, Morgantown, WV 26506, USA}
\affiliation{Center for Gravitational Waves and Cosmology, West Virginia University, Chestnut Ridge Research Building, Morgantown, WV 26505, USA}
\author[0000-0001-8800-0192]{Caner Unal}
\affiliation{Department of Physics, Ben-Gurion University of the Negev, Be'er Sheva 84105, Israel}
\affiliation{Feza Gursey Institute, Bogazici University, Kandilli, 34684, Istanbul, Turkey}
\author[0000-0002-4162-0033]{Michele Vallisneri}
\affiliation{Jet Propulsion Laboratory, California Institute of Technology, 4800 Oak Grove Drive, Pasadena, CA 91109, USA}
\affiliation{Division of Physics, Mathematics, and Astronomy, California Institute of Technology, Pasadena, CA 91125, USA}
\author[0000-0003-4700-9072]{Sarah J. Vigeland}
\affiliation{Center for Gravitation, Cosmology and Astrophysics, Department of Physics, University of Wisconsin-Milwaukee,\\ P.O. Box 413, Milwaukee, WI 53201, USA}
\author[0000-0001-9678-0299]{Haley M. Wahl}
\affiliation{Department of Physics and Astronomy, West Virginia University, P.O. Box 6315, Morgantown, WV 26506, USA}
\affiliation{Center for Gravitational Waves and Cosmology, West Virginia University, Chestnut Ridge Research Building, Morgantown, WV 26505, USA}
\author[0000-0002-6020-9274]{Caitlin A. Witt}
\affiliation{Center for Interdisciplinary Exploration and Research in Astrophysics (CIERA), Northwestern University, Evanston, IL 60208}
\affiliation{Adler Planetarium, 1300 S. DuSable Lake Shore Dr., Chicago, IL 60605, USA}
\author[0000-0002-0883-0688]{Olivia Young}
\affiliation{School of Physics and Astronomy, Rochester Institute of Technology, Rochester, NY 14623, USA}
\affiliation{Laboratory for Multiwavelength Astrophysics, Rochester Institute of Technology, Rochester, NY 14623, USA}

\collaboration{1000}{The NANOGrav Collaboration}
\noaffiliation
\correspondingauthor{The NANOGrav Collaboration}
\email{comments@nanograv.org}

\begin{abstract}
Recently we found compelling evidence for a gravitational wave background with Hellings and Downs (HD) correlations in our 15-year data set. These correlations describe gravitational waves as predicted by general relativity, which has two transverse polarization modes. However, more general metric theories of gravity can have additional polarization modes which produce different interpulsar correlations. In this work we search the NANOGrav 15-year data set for evidence of a gravitational wave background with quadrupolar Hellings and Downs (HD) and Scalar Transverse (ST) correlations. We find that HD correlations are the best fit to the data, and no significant evidence in favor of ST correlations. While Bayes factors show strong evidence for a correlated signal, the data does not strongly prefer either correlation signature, with Bayes factors $\sim 2$ when comparing HD to ST correlations, and $\sim 1$  for HD plus ST correlations to HD correlations alone. 
However, when modeled alongside HD correlations, the amplitude and spectral index  posteriors for ST correlations are uninformative, with the HD process accounting for the vast majority of the total signal. Using the optimal statistic, a frequentist technique that focuses on the pulsar-pair cross-correlations, we find median 
signal-to-noise-ratios of 5.0 for HD and 4.6 for ST correlations when fit for separately, and median signal-to-noise-ratios of 3.5 for HD and 3.0 for ST correlations when fit for simultaneously. While the signal-to-noise-ratios for each of the correlations are comparable, the estimated amplitude and spectral index for HD are a significantly better fit to the total signal, in agreement with our Bayesian analysis.

\end{abstract}


\section{\label{sec:intro}Introduction}

Einstein's theory of general relativity (GR) predicts the existence of gravitational waves (GWs) with two transverse polarization modes that propagate at the speed of light \citep{Eardley:1973zuo,Eardley2}. Observations by the LIGO collaboration have shown that GR best describes gravitational radiation from massive freely accelerating objects in the universe \citep{LIGO_firstDet, LIGO_Virgo_recentDet}. However, pulsars and pulsar timing array (PTA) experiments offer a unique opportunity to probe other possible metric theories of gravity external to Einstein's GR.

Modified theories of gravity are often introduced to resolve some of the current challenges facing fundamental physics, such as the nature of dark matter, dark energy, and in attempts to reconcile quantum mechanics and gravity (see, e.g., \citealt{berti_2015} and references therein). In metric theories of gravity, there can be up to six possible GW polarization modes \citep{Eardley:1973zuo,Eardley2,Will:1993hxu}. PTA searches for non-Einsteinian polarization modes may provide evidence for modified gravity theories by uncovering the different correlation patterns associated with such modes \citep{Chamberlin:2011ev,Yunes:2013dva,Gair:2015hra,Cornish:2017oic,NG15new_physics}. 

Millisecond pulsars (MSPs) emit radio beams from their magnetic poles and are extremely stable rotators. They appear to us as point sources of periodic radio bursts that arrive on Earth with a consistency that rivals that of atomic clocks \citep{Matsakis:1997,Hobbs:2012,Hobbs:2020a}. Pulsar timing experiments exploit the regularity of MSPs to search for low-frequency ($\sim $1-100 nHz) GWs by measuring deviations from the expected arrival time of radio pulses \citep{Sazhin:1978, Detweiler:1979wn}. Moreover, an array of these MSPs allows us to search for correlations between deviations of times of arrivals (TOAs) of pulses from MSP pairs \citep{hd83, FB:1990}. 

The North American Nanohertz Observatory for Gravitational Waves (NANOGrav), the European Pulsar Timing Array (EPTA), the Chinese Pulsar Timing Array (CPTA), and the Parkes Pulsar Timing Array (PPTA) are the PTAs that possess the most sensitive datasets capable of measuring nHz GWs. NANOGrav, the EPTA, and the PPTA have seen strong evidence for a common red noise process \citep{12gwb, EPTA_DR2_Noise, PPTA_DR2_gwb}. Most recently in \citet{15yrGWB} (hereafter referred to as \citetalias{15yrGWB}), NANOGrav has found compelling evidence for quadrupolar correlations \citep{hd83}, while the EPTA and PPTA have seen these correlations at varying levels of significance \citep{Antoniadis:2023rey, Reardon:2023gzh}. In this paper, we complement our work in \citetalias{15yrGWB} by searching for evidence for Scalar Transverse correlations from the non-Einsteinian breathing polarization mode of gravity. Previous work in \citep{12altpol} has shown preference for a Scalar-Transverse (ST) and GW-like monopolar correlations. However, these correlations were not significant as they were not robust to the solar system ephemeris and were associated with pulsar J0030$+$0451. In \S\ref{sec:background}, we review the theoretical background required to identify and search for a general transverse polarization mode of gravity using PTAs. In \S\ref{sec:analyses}, we then describe the analyses performed, both using Bayesian and frequentist approaches. Lastly, in \S\ref{sec:conclusion}, we present the evidence for/against the existence of ST correlations.

\section{\label{sec:background}Theoretical Background}
In this section, we will first review the basics of gravitational wave polarization modes, \S\ref{subsec:Generalized Polarization Modes in Metric Theories of Gravity}, and proceed to outline the theoretical considerations needed to predict the signature of such modes in a PTA GWB signal in 
\S\ref{subsec:Pulsar Timing with an Isotropic Gravitational Wave Background}. Finally, in \S\ref{subsec:Restriction to Transverse Modes}, we will explicitly describe the model for a general transverse GWB signal which we will later search for in \S\ref{sec:analyses} using the NANOGrav 15-year data set.
\subsection{\label{subsec:Generalized Polarization Modes in Metric Theories of Gravity}Generalized Polarization Modes in Metric Theories of Gravity}
\label{subsec:polarization modes}
In metric theories of gravity, there can be between two and six independent polarization modes for GWs \citep{Eardley:1973zuo}.
These modes are the ``electric" components of the Riemann tensor $R_{0i0j}$, where $i$ and $j$ are the spatial components.
These components were originally found by \citet{Newman:1961qr} making use of tetrad and spinor calculus.

For the purposes of this work, we assume a coordinate system such that a null plane-GW travels along the $+z$-axis at the speed of light ($c$), where the components of the Riemann tensor only depend on the retarded time $u=t-z/c$.
The assumptions lead to the following coefficients which depend on combinations of the independent electric components of the Riemann tensor:
\begin{subequations}
    \begin{align}
        \Psi_2(u) &= -\frac{1}{6} R_{0303}(u) \,, \\
        \Psi_3(u) &= -\frac{1}{2}R_{0103} + \frac{i}{2}R_{0203} , \\
        \Psi_4(u) &= -R_{0101} + R_{0202} + 2iR_{0102} \,, \\
        \Phi_{22}(u) &= -R_{0101} - R_{0202} \,.
    \end{align}
\end{subequations}
We may relate these to a matrix of the GW polarization modes by 
\begin{equation}
\label{eq:PolarizationMatrix}
    S^{ij} =
        \begin{bmatrix}
            A_B + A_+ & A_\times & A_{V1} \\
            A_\times & A_B - A_+ & A_{V2} \\ 
            A_{V1} & A_{V2} & A_L
        \end{bmatrix} \,,
\end{equation}
where
\begin{subequations}
    \begin{align}
        A_+ &= {\rm Re}(\Psi_4) \,, \\
        A_\times &= {\rm Im}(\Psi_4) \,, \\
        A_B &= \Phi_{22} \,, \\
        A_{V1} &= {\rm Re}(\Psi_3) \,, \\
        A_{V2} &= {\rm Im}(\Psi_3) \,, \\
        A_L &= \Psi_2 \,.
    \end{align}
\end{subequations}
Here, $A_+$ and $A_\times$ represent the two tensor modes of GWs, the only two allowed by GR.
The shear modes are given by $A_{V1}$ and $A_{V2}$, while the scalar breathing and scalar longitudinal modes are $A_B$ and $A_L$ respectively.
Searching for the coefficients in Eq.~\eqref{eq:PolarizationMatrix} allows for a theory-independent way to perform a test of gravity, without the need to be concerned with the specifics of any metric theory of gravity. We will utilize this technique in searches for a GWB using PTAs.

\subsection{\label{subsec:Pulsar Timing with an Isotropic Gravitational Wave Background}Pulsar Timing and Isotropic Gravitational Wave Background}
A GW propagating through the Earth-pulsar line of sight will induce a change in the expected time of arrival for the pulsar's radio pulse. These perturbations were first calculated in the late 1970s \citep{Sazhin:1978, Detweiler:1979wn} and have since been used to predict the GWB signature. For pulsar timing, the measured variation in the pulse TOAs can be used to calculate GW induced residuals, $R_a^{\rm GW}$, of pulsar $a$ following the relation
\begin{equation}
\label{eq:Residuals}
    R_a^{\rm GW} = \int_0^t dt^\prime z_a(t^\prime),
\end{equation}
which is the quantity measured directly in PTAs, where $z_a$ is the GW induced redshift. For a detailed explanation of Eq.~\eqref{eq:Residuals}, refer to \citet{12altpol} and \citet{Chamberlin:2011ev}.

The fractional energy density of the background is given to be 
\begin{equation}
    \Omega_{\rm GW}(f) = \frac{1}{\rho_c} \frac{d \rho_{\rm GW}}{d \ln(f)} 
\end{equation}
where $\rho_{\rm GW}$ is the energy of the gravitational wave, $f$ is the frequency, and $\rho_c$ is the critical density necessary for a closed universe.
For the purposes of this analysis, we assume the GWB is produced by a large number of independent, weak, unresolvable sources isotropically distributed throughout the sky.
Hence, the correlation between the strain functions is written as 
\begin{equation}
\label{eq:Correlation1}
    \langle \tilde{h}_g(f,\hat{\Omega}) \tilde{h}_{g^\prime}^* (f^\prime,\hat{\Omega}^\prime) \rangle = \frac{\delta (f-f^\prime)}{2} \frac{\delta^2 ( \hat{\Omega},\hat{\Omega}^\prime )}{4 \pi} \frac{\delta_{gg^\prime}}{2} S_h(f)\,,
\end{equation}
where $S_h(f)$ is the one-sided power-spectral-density of the GWB; related to $\Omega_{\mathrm GW}(f)$ by
\begin{equation}
    \Omega_{\rm GW}(f) = \frac{2 \pi^2}{3 H_0^2}f^3 S_h(f)\,.
\end{equation}

The spectral characteristics of the GWB are often described via the \emph{characteristic strain}
\begin{equation}
\label{eq:Strain1}
    h_{c}(f) = \sqrt{f S_h(f)}\,.
\end{equation}
This quantity is useful, as it includes the effects of the number of cycles during the GW source in-spiral throughout the frequency band $\sqrt{f}$ as discussed in~\citet{taylor2021nanohertz}.
While several models exist for describing the nature of $h_c(f)$ \citepalias{15yrGWB}, in this work, we will restrict ourselves to that of a power-law model for each polarization $g$  such that
\begin{equation}
\label{eq:Strain2}
    h_{g}(f) = A_g \left( \frac{f}{f_{\rm yr}} \right)^{\alpha_g} \,
\end{equation}
where $A_g$ is a dimensionless amplitude, $f_{\rm yr}$ is the reference frequency, and $\alpha_{g}$ is the spectral index.
Using Eqs.~\eqref{eq:Correlation1},~\eqref{eq:Strain1}, and~\eqref{eq:Strain2}, we can find the cross-correlation estimator between pulsars $a$ and $b$ \citep{Chamberlin:2011ev}:
\begin{align}
\langle R_a^{\rm GW}R_b^{\rm GW}\rangle &= \int_{f_L} ^{f_H} df \{S_{ab}(f)\}, \\
\begin{split}
    \label{eq:CrossCorrelationSpectralDensity1}
    S_{ab}(f) &= \frac{1}{8 \pi^2 f^3 }\sum_g \Gamma^g_{ab}(f) h_{g}^2(f)\\ 
    &= \frac{3}{2} \sum_g P_{g} \Gamma^g_{ab}(f) \ ,
\end{split}
\end{align}
where $f_L$ and $f_H$ are lower and upper frequencies and $\Gamma_{ab}^g(f)$ is the \emph{overlap reduction function} (ORF) which is related to the spatial geometry of the two pulsars in relation to the Earth and we have introduce $P_{g}$ defined as 
\begin{align}
\label{eq:gwbpower}
    P_{g}(f) &= \frac{ A_{g}^2}{12 \pi^2 f^3 } \left( \frac{f}{f_{yr}}\right)^{3-\gamma_{g}}. \nonumber \\ \,
\end{align}
In the above, to align with the more widely used terminology for spectral index, we have made the reparametrization $\gamma_g = 3 - 2\alpha_g$.

\subsection{\label{subsec:Restriction to Transverse Modes}Restriction to Transverse Modes}
As discussed in \S\ref{subsec:polarization modes}, there exist between two and six possible independent polarization modes for a GW in metric theories of gravity.
Calculating the effects of longitudinal modes requires additional steps and assessments such as accurate knowledge of distances to the pulsars, handling the frequency dependence of the ORF, as well as having a significant number of pulsars at small-angular separations to capture the unique ORF signature of such polarization modes \citep{12altpol}. Thus, we will restrict ourselves to the three transverse modes, $A_+$ $A_\times$, and $A_B$, for the purposes of this paper.

Given only the transverse tensor and scalar modes, we may generalize Eq.~\eqref{eq:CrossCorrelationSpectralDensity1} \citep{firstalt, 12altpol} as
\begin{align}
\label{eq:CrossCorrelationSpectralDensity2}
    S_{ab}(f) = \frac{3}{2} \left( P_{\rm TT} \Gamma^{\rm TT}_{ab}+ P_{\rm ST} \Gamma^{\rm ST}_{ab} \right). \,
\end{align} 
It is worth pointing out that the effect of dipole radiation of binary sources in non-GR metric theories of gravity \citep{firstalt}, is accounted for by treating the spectral index $\gamma_g$ as a free parameter in our statistical models.

The ORFs for the TT and ST modes have been calculated previously in \citet{Chamberlin:2011ev} and \citet{Gair:2015hra}, 
\begin{subequations}
    \begin{align}
        \Gamma^{\rm TT}_{ab} &= \frac{\delta_{ab}}{2} + \frac{3}{2} \left( \frac{1}{3} + k_{ab} \left( \ln{k_{ab}} - \frac{1}{6}\right)\right) \,, \\
        \Gamma^{\rm ST}_{ab} &= \frac{\delta_{ab}}{2} + \frac{1}{8} \left( 3 + \cos{\xi_{ab}}\right)\,,
    \end{align}
\end{subequations}
with $\xi_{ab}$ being the angular distance on the sky between pulsars $a$ and $b$ and 
\begin{equation}
    k_{ab} = \frac{1}{2} \left( 1 - \cos{\xi_{ab}} \right)\,.
\end{equation}
\begin{figure}
    \centering
    \includegraphics[width = \linewidth]{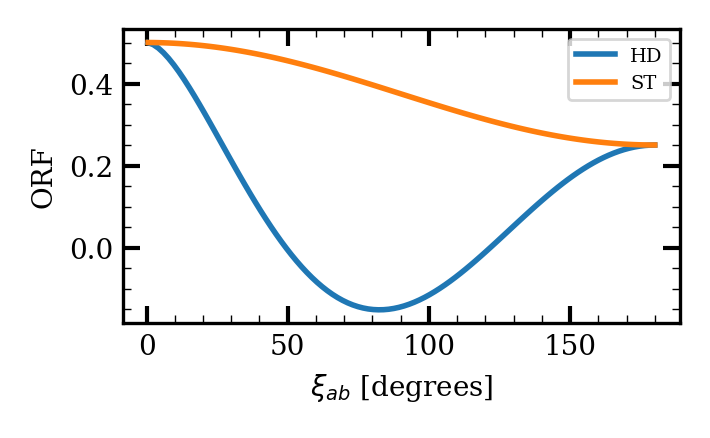}
    \caption{A plot of transverse ORFs as a function of angular separation. The blue curve describes the Hellings and Downs curve, which is produced by the TT polarization mode, while the orange curve describes the shape of the correlations induced by the ST polarization mode of gravity. }
    \label{fig:ORFs}
\end{figure}
A plot of the transverse ORFs as a function of angular separation is shown in \autoref{fig:ORFs}. Where $\Gamma^{\rm TT}_{ab}$ is represented by the more widely known Hellings and Downs (HD) curve \citep{hd83}, and henceforth the TT mode will be represented by HD.
With this structure in hand, we may proceed to the analysis of the NG15yr data and the investigation of the general transverse modes.

\section{\label{sec:analyses}Searches for a General Transverse GWB in the NANOGrav 15 Year Data Set}

In this section, we complement our previous work in \citetalias{15yrGWB} by analyzing the NANOGrav 15 year data set for statistical significance of HD plus ST correlations. We first describe the pulsar noise modeling in \S\ref{subsec:modeling details}, and then we present the results of the Bayesian and frequentist analyses in \S\ref{subsec:Bayesian} and \S\ref{subsec:os} respectively. We will take an agnostic approach to the mixing between the HD and the ST polarization modes of gravity by allowing each mode to possess its own independent power-spectral-density as suggested by Eq.~\eqref{eq:CrossCorrelationSpectralDensity2}.

\subsection{\label{subsec:modeling details}Noise Modeling Details}
Through individual pulsar analyses we obtain posteriors for both the red and white noises intrinsic to each pulsar. In all of the analyses, these intrinsic noises are modeled as a power-law with varying spectral index as well as varying amplitudes. We model the power spectra using frequency bins from $1/T_{\text{obs}}$ to $30/T_{\text{obs}}$ for $T_{\text{obs}}$ being the longest observational baseline among the considered pulsars in the data set. The white noise is described by three parameters: a linear scaling of TOA uncertainties, noise added in quadrature to the TOA uncertainties, and noise common to a given epoch at all frequency subbands. These parameters are called EFAC, EQUAD, and ECORR, respectively and are set to their fixed values in \citetalias{15yrGWB}. For detailed explanations of these parameters, refer to \citep{NG15yrNoise}. 

The common red noise process is modeled as a power-law model in three ways: CURN, HD, and ST, all of which are modeled using frequency bins from $1/T_{\text{obs}}$ to $14/T_{\text{obs}}$. Adapted from the naming convention in \citetalias{15yrGWB}, $\mathrm{CURN}$ refers to the modeling of the red noise common process as an uncorrelated process (i.e., $\Gamma_{ab}^{\text{CURN}} = \delta_{ab}$). Whereas HD and ST refer to modeling of the red noise common process as a correlated process with HD and ST curves as their respective correlation signatures.

The upper and lower bounds of the model parameters we use are shown below. Note that the subscript ``$\text{int}$" refers to the intrinsic red noise processes while the subscript $g$ refers to the common red noise process (i.e., CURN, HD, or ST). 
\begin{align}
\begin{split}
    log_{10} A_{\text{int}} &\sim \text{Uniform}(-20, -11), \\
    log_{10}A_{\text{g}}&\sim \text{Uniform}(-18, -11),\\ 
    \gamma_{\text{int}}, \gamma_{g} &\sim \text{Uniform}(0, 7).
\end{split}
\end{align}
Refer to \citetalias{15yrGWB} and \cite{NG15yrNoise} for a more detailed explanation of the noise modeling adapted for the analyses of the NANOGrav 15 year data set.
\subsection{\label{subsec:Bayesian}Bayesian Analyses}

\begin{figure}[!b]
    \centering
    \includegraphics[width = \linewidth]{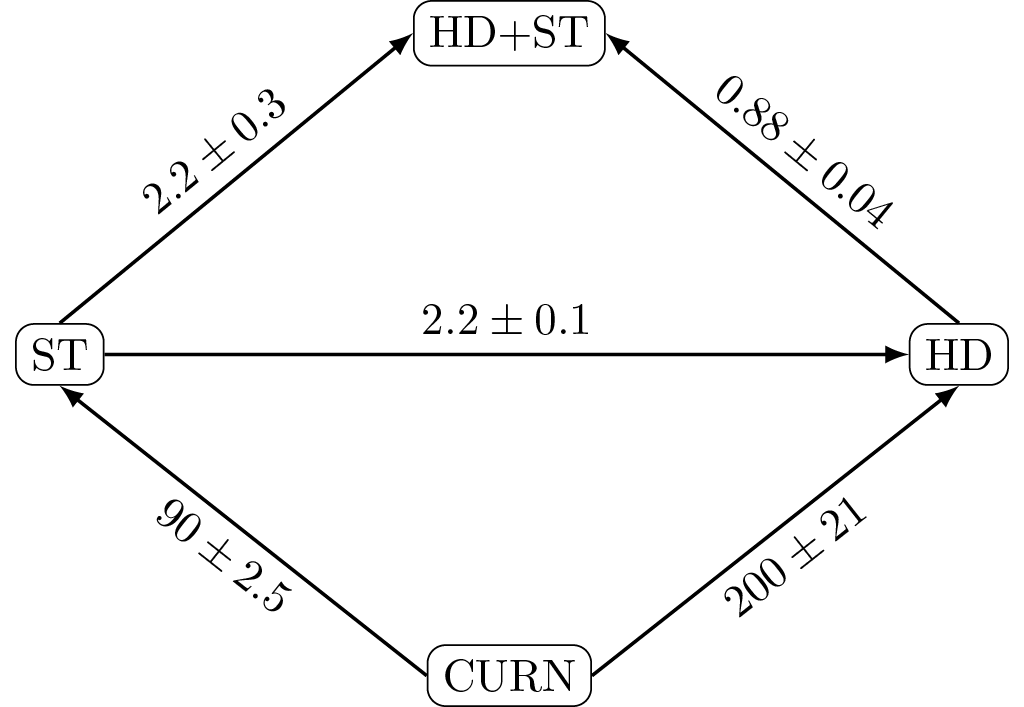}
    \caption{Bayes factors for various model comparisons between ST, HD, and CURN. Overall, the HD model is preferred over CURN and ST. Modeling ST alongside HD give about equal odds over HD and ST only. All model comparisons are agnostic with respect to the spectral index of each model. See~\S\ref{subsec:modeling details} for more details. The uncertainties are estimated using bootstrapping and Markov model techniques of \protect\citet{Heck_uncertainty}. All Bayes factors are presented as the model at the end of an arrow over the starting model of an arrow. For example, for the arrow pointing from CURN to ST, the values are BFs for ${ST} \divslash {CURN}$.}
    \label{figure:bayesogram}
\end{figure}
    
\begin{figure*}[!ht]
\centering
\subfloat[]{%
\includegraphics[width =0.49\linewidth]{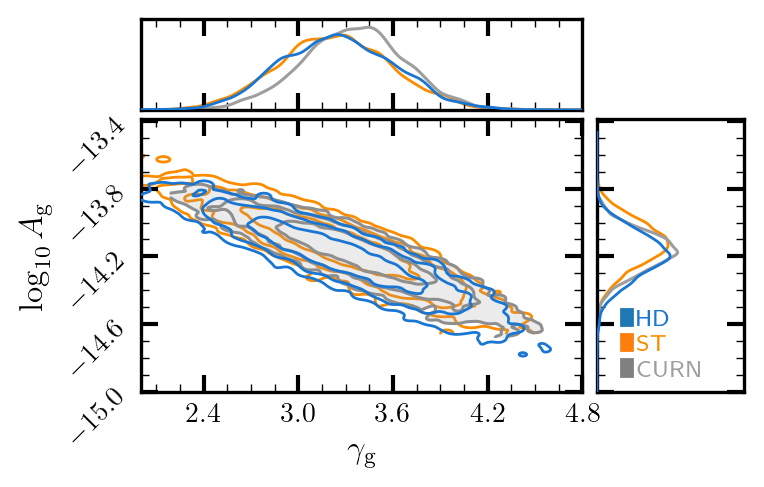}%
}\quad
\subfloat[]{%
\includegraphics[width =0.49\linewidth]{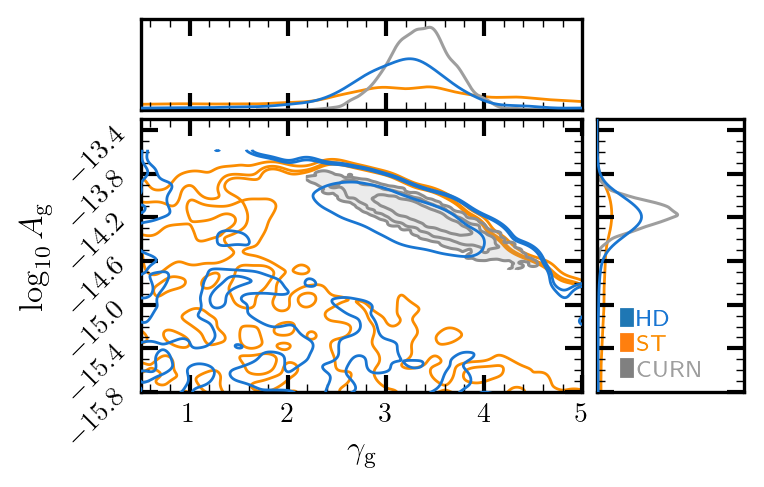}%
}
\caption{(a) Bayesian probability posterior distributions of $\log_{10}A_g$ and $\gamma_g$ from a HD correlated model ($\color{blue}blue$) and a ST correlated model ($\color{orange}orange$) showing the 1$\sigma$/2$\sigma$/3$\sigma$ credible regions. Plotted for comparison is CURN ($\color{gray}grey$) posterior distributions for $\log_{10}A_g$ and $\gamma_g$ parameters. Each correlated signal is able to explain the total signal. (b) Bayesian probability posterior distributions of $\log_{10}A_g$ and $\gamma_g$ for HD ($\color{blue}blue$) and ST ($\color{orange}orange$) from a HD+ST correlated model showing 1$\sigma$/2$\sigma$/3$\sigma$ credible regions. Plotted for comparison is CURN ($\color{gray}grey$) posterior distributions for $\log_{10}A_g$ and $\gamma_g$ parameters. The plots suggest that the posterior distribution for ST is uninformative and HD adequately describes the total signal recovered by CURN.}

\label{fig:bayes_amps1}
\label{fig:bayes_amps}
\end{figure*}

Our Bayesian analyses follow \S\ref{subsec:modeling details} as well as \citetalias{15yrGWB}. In short, in terms of a likelihood function, all the various noise modelings follow from \citep{NG15code}
\begin{equation}
    \label{eq:fast_like}
    p(\vect{\delta t} \mid \vect{\eta}) = \frac{1}{\sqrt{\det(2 \pi \vect{K})}} \exp\left(-\frac{1}{2}\vect{\delta t}^T \vect{K}^{-1}\vect{\delta t}\right)\,,
\end{equation}
where
\begin{equation}
    \vect{K}=\vect{D}+\vect{F} \vect{\phi} \vect{F}^T,
\end{equation}
and we then use the Woodbury matrix identity to invert this covariance matrix. 
We find
\begin{equation}
    \vect{K}^{-1}=\vect{D}^{-1}-\vect{D}^{-1} \vect{F} \vect{\Theta} \vect{F}^T \vect{D}^{-1}\,,
\end{equation}
with
\begin{equation}
    \label{eq:phi_inv}
    \vect{\Theta}=\left(\vect{\phi}^{-1}+\vect{F}^T \vect{D}^{-1} \vect{F}\right)^{-1}\,.
\end{equation}
In the above, $\vect{F}$ is a matrix with alternating columns of sine and cosine components representing a discrete Fourier transform of the red noise processes, $\vect{D}$ is covariance matrix for the white noise parameters, $\vect{\phi}$ is the covariance matrix of the red noise components. 

We use Bayesian analyses to compare several models of interest via Bayes factor estimation, \autoref{figure:bayesogram}, and to obtain posterior distributions for $\log_{10}A_g$ and $\gamma_g$ for HD and ST signals, \autoref{fig:bayes_amps}. 
In \autoref{figure:bayesogram}, we observe that correlated Bayesian models are preferred over the uncorrelated model. The most favored model is a GWB with HD correlations with a Bayes factor of 200.   
When ST is modeled alongside HD, Bayes factors are uninformative given they are on the order of unity when compared to each correlation alone. 

We can use the transitive nature of Bayes factors as a consistency check of our results. For instance, going around the bottom half of \autoref{figure:bayesogram} we can take the Bayes factor of $\rm{ST} \divslash \rm{CURN}$ and multiply by the Bayes factor of $\rm{HD} \divslash \rm{ST}$ to obtain the Bayes factor of $\rm{HD} \divslash \rm{CURN}$. This results in $90 \times 2.2  = 198$, which is consistent with the Bayes factor for $\rm{HD} \divslash \rm{CURN}$ of $\sim 200$ we obtained by directly comparing those two models. 

We note that in \autoref{fig:bayes_amps} when fitting for one correlation signature that both HD correlations and ST correlations are able to explain the total signal. This agrees with the large Bayes factors favoring these models over CURN. However, the recovered power spectral estimates for ST are poor when modeled alongside HD. To check the consistency of the power spectral estimates we see that 
$\log_{10}A_{\rm{CURN}} = -14.17^{+0.12}_{-0.13}$
and
$\gamma_{\rm{CURN}} = 3.35^{+0.32}_{-0.32}$ 
(median values with 68\% credible interval).
The ST values are 
$\log_{10}A_{\rm{ST}} = 15.03^{+0.87}_{-1.92}$ 
and 
$\gamma_{\rm{ST}} = 3.33^{+1.53}_{-1.53}$, 
while the HD values are $\log_{10}A_{\rm{HD}} = 14.24^{+0.18}_{-0.56}$ 
and 
$\gamma_{\rm{HD}} = 3.17^{+0.51}_{-0.61}$. 
We see that values for CURN and HD are more consistent with each other. While the ST spectral estimates do overlap with the median of the CURN spectral estimates, we observe the 68\% credible region for $\gamma_{\rm{ST}}$ and $\log_{10}A_{\rm{ST}}$ expand over about 43\% and 31\% of the prior region, respectively. Therefore, the addition of the ST correlation yields no additional information and we see that the HD signal in this model explains most of the total signal.

\subsection{\label{subsec:os}Optimal Statistic Analyses}
\begin{figure*}[!htp]
    \centering
    \includegraphics[width =\linewidth]{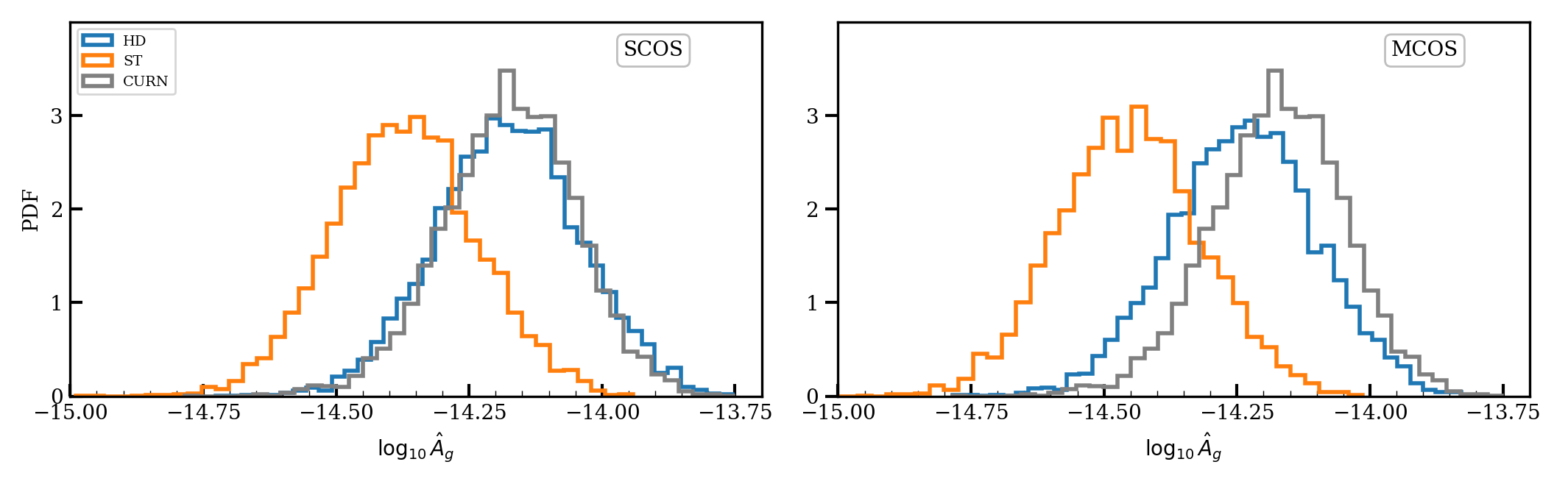}
    \caption{Distributions of the recovered amplitudes from single-component noise marginalized optimal statistic (SCOS)(\emph{left}) and multi-component noise marginalized optimal statistic (MCOS)(\emph{right}) for HD ($\color{blue}blue$) and ST ($\color{orange}orange$) correlations. Additionally, CURN ($\color{gray}grey$) is plotted for comparison to determine the consistency with the common red noise process. 
}
    \label{fig:OSamps}
\end{figure*}

\begin{figure}[!htp]
    \centering
    \includegraphics[width =\linewidth]{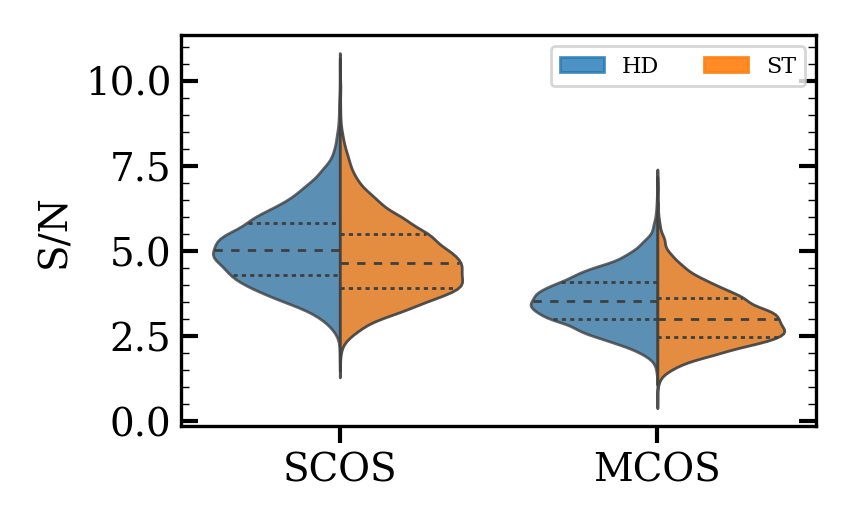}
    \caption{Distributions of the S/N for HD ($\color{blue}blue$) and ST ($\color{orange}orange$) correlations from the multi-component (MCOS) and single component (SCOS) noise marginalized optimal statistic techniques. The dashed lines correspond to the median values and the first and third quartiles. The median S/N values for HD are greater than ST, but are within their inter-quartile range. }
    \label{fig:SNR}
\end{figure}

The optimal statistic \citep{OPTSTAT0, TimeDomainPaper} allows for a robust and computationally inexpensive analysis of the correlation content of a PTA data set. The amplitude and the uncertainty of the pair-wise cross correlations are estimated by maximizing the ratio of the likelihood of the fiducial GWB over the \emph{noise-only} model. The fiducial model contains a GWB signal along with intrinsic red and white noise components while the noise-only model includes the intrinsic noises and a common uncorrelated red noise process. We have employed the \emph{noise-marginalized} version of the optimal statistic technique in which $10^{4}$ random draws from the posteriors of all of the model parameters of a $\text{CURN}$ model are used to estimate the required power spectra. Additionally, since our goal is to search for a general transverse GWB signal in which two non-orthogonal types of correlations might simultaneously exist in the data set, a chi-squared statistic fitting for both HD and ST correlations is used to find the optimal estimators of the \emph{signal-to-noise-ratio} (S/N or SNR) and the amplitude of the correlated signal ($\hat{A}_g$) for each polarization mode. Note that $\hat{A}_g$ differs from ${A}_g$ as the former is an optimal estimator of the latter. See \citet{NMOS} and \citet{MCOS} for more details.

The estimated amplitudes for single component (SCOS) and multi-component (MCOS) noise marginalized optimal statistics are shown in \autoref{fig:OSamps}. We see that for SCOS that the amplitude reconstruction of the HD correlations is in excellent agreement with the CURN amplitude posterior, whereas the estimated amplitude for ST correlations is only consistent with CURN. For MCOS the correlations are fit for simultaneously so the total power is split and the fit correlations are shifted towards smaller amplitude values and less consistent with CURN. However, the HD correlations explain most of the total CURN signal when both correlations are present. 

The SNR for SCOS and MCOS are shown in \autoref{fig:SNR}. We note that the median SNRs for HD correlations, 5.0 and 3.5 for SCOS and MCOS, respectively, are larger than the median SNRs for ST correlations, 4.6 and 3.0 for SCOS and MCOS, respectively. The difference in median SNR values are not significant as the medians lie within the inter-quartile ranges of each other. While the SNR values are similar for HD and ST correlations, as noted before, the consistency of the estimated HD amplitude with CURN suggest that HD, not ST, correlations make up most of the common red noise process.

\section{\label{sec:conclusion}Discussion} 
NANOGrav's 15-year data set shows compelling evidence for quadrupolar HD inter-pulsar correlations. In this work, we explored the possibility of deviations from the HD curve caused by the presence of an additional scalar-transverse (ST) mode.

Our Bayesian analyses show the Bayes factor for HD over ST is $\sim 2$, and the Bayes factor for a model with both correlations compared to a model with just HD is $\sim 1$. Taking the spectral parameter recovery into account, as in \autoref{fig:bayes_amps}, we found each correlation, when fit for individually, is in agreement with CURN. We also found more informative $\log_{10}A_g$ and $\gamma_g$ recovery for HD than ST, and HD parameters show better agreement with CURN spectral parameters when correlations are included together. While these analyses do not rule out the possibility of ST correlations in our data, they show there is no statistical need for an additional stochastic process with ST correlations. 

This is also the case for our frequentist analyses. When fitting the interpulsar correlation data for a single correlation signature, we find that HD correlations completely account for the total signal due to the amplitudes consistency with CURN, but ST correlations are only somewhat consistent. When we fit for both correlations simultaneously, we still see that HD correlations are able to explain most of the total signal. For the SNR, we find that the median values for HD correlations are larger than ST correlations, but are similar and lie within inter-quartile ranges. 

Even though we cannot fully rule out ST correlations, Einsteinian polarization modes with HD correlations are present in all metric theories of gravity. Thus, even though we do not find a convincing Bayes factor favoring HD correlations over ST correlations and they have similar SNR values in our frequentist analyses, there is no metric theory that predicts only ST GWs. In addition, we no longer report higher SNR and Bayes factors for ST correlations as we did in \citet{12altpol}. We have seen a larger increases in favor of HD correlations than ST correlations in both SNR and Bayes factors. These changes are consistent with simulations in \citet{12altpol} and, with no evidence indicating otherwise, we expect this trend to continue with additional data.
 
We also performed dropout analysis tests, similar to what was done in \citet{12altpol}, to determine if particular pulsars play a role in the observed ST significance. We found that J0030$+$0451 and J0613$-$0200 are responsible for a majority of the ST significance. When we remove these two pulsars from the analysis, we find that the Bayes factor  
for $\rm{HD} \divslash \rm{CURN}$ increases to $\sim 600$, while the Bayes factor for $\rm{ST} \divslash \rm{CURN}$ is reduced to $\sim 30$. We suspect 
improved noise modeling (as used in \cite{IPTA2_CWNoise} and \cite{NG15yrNoise}) on these and other pulsars will shed some light on this, and we leave this for future work.

Other recent work \citep{allen_hdvariance,AllenJoe_variance} has shown that the HD correlation signature has a cosmic variance. This idea is not addressed within this manuscript, but introduces increasing complexity in detecting alternative polarizations of gravity. An alternative polarization mode now not only needs to be disentangled from the HD correlations, but also the variance of the HD correlations to be detected. Impacts of these effects will be addressed in future work.

Future prospects for performing tests of gravity using PTA data are compelling. Large observational baselines as well as the addition of more millisecond pulsars to the observing array will enable more robust and sensitive searches for additional GW polarization modes. In this work, we reported on one test of gravity in which we searched for evidence for a scalar transverse polarization mode. While we did not find substantial evidence for or against this mode, the situation may change in the future due to the nature of the PTA data sets. It is also important to note that a  number of the observed pulsars are dominated by white and intrinsic red noise process which could be suppressing a GW-sourced signal. For the case of a GWB, as we obtain more data on these pulsars, we will be able to provide more definitive answers about the possibility of the existence or absence of additional polarization modes of gravity.

\section{\label{subsec:acknowledgments} Acknowledgments}

\textit{Author contributions: }
An alphabetical-order author list was used for this paper in recognition of the fact that a large, decade timescale project such as NANOGrav is necessarily the result of the work of many people. All authors contributed to the activities of the NANOGrav collaboration leading to the work presented here, and reviewed the manuscript, text, and figures prior to the paper's submission. 
Additional specific contributions to this paper are as follows.
%
G.A., A.A., A.M.A., Z.A., P.T.B., P.R.B., H.T.C., K.C., M.E.D., P.B.D., T.D., E.C.F., W.F., E.F., G.E.F., N.G., P.A.G., J.G., D.C.G., J.S.H., R.J.J., M.L.J., D.L.K., M.K., M.T.L., D.R.L., J.L., R.S.L., A.M., M.A.M., N.M., B.W.M., C.N., D.J.N., T.T.P., B.B.P.P., N.S.P., H.A.R., S.M.R., P.S.R., A.S., C.S., B.J.S., I.H.S., K.S., A.S., J.K.S., and H.M.W. through a combination of observations, arrival time calculations, data checks and refinements, and timing model development and analysis; additional specific contributions to the data set are summarized in \citet{NG15data}.
%
D.M.D led the search and coordinated the writing of the paper. D.M.D, N.L., A.S., S.C.S., and J.A.T. performed the Bayesian and frequentist analysis. J.B. and K.D.O cross-checked the Bayes factor values. B.B., J.S.H., K.D.O., X.S., J.P.S., S.R.T., and S.J.V. provided feedback and guidance on searches and analysis. N.L. and X.S. provided first insights to dropping J0030+0451 and J0613-0200. D.M.D, N.L., A.S., S.C.S., and J.A.T. wrote the paper.

\textit{Acknowledgments: }The work contained herein has been carried out by the NANOGrav collaboration, which receives support from the National Science Foundation (NSF) Physics Frontier Center award numbers 1430284 and 2020265, the Gordon and Betty Moore Foundation, NSF AccelNet award number 2114721, an NSERC Discovery Grant, and CIFAR. The Arecibo Observatory is a facility of the NSF operated under cooperative agreement (AST-1744119) by the University of Central Florida (UCF) in alliance with Universidad Ana G. M$\acute{\text{e}}$ndez (UAGM) and Yang Enterprises (YEI), Inc. The Green Bank Observatory is a facility of the NSF operated under cooperative agreement by Associated Universities, Inc.
L.B. acknowledges support from the National Science Foundation under award AST-1909933 and from the Research Corporation for Science Advancement under Cottrell Scholar Award No. 27553.
P.R.B. is supported by the Science and Technology Facilities Council, grant number ST/W000946/1.
S.B. gratefully acknowledges the support of a Sloan Fellowship, and the support of NSF under award \#1815664.
The work of R.B., R.C., D.D., N.La., X.S., J.P.S., and J.A.T. is partly supported by the George and Hannah Bolinger Memorial Fund in the College of Science at Oregon State University.
M.C., P.P., and S.R.T. acknowledge support from NSF AST-2007993.
M.C. and N.S.P. were supported by the Vanderbilt Initiative in Data Intensive Astrophysics (VIDA) Fellowship.
K.Ch., A.D.J., and M.V. acknowledge support from the Caltech and Jet Propulsion Laboratory President's and Director's Research and Development Fund.
K.Ch. and A.D.J. acknowledge support from the Sloan Foundation.
Support for this work was provided by the NSF through the Grote Reber Fellowship Program administered by Associated Universities, Inc./National Radio Astronomy Observatory.
Support for H.T.C. is provided by NASA through the NASA Hubble Fellowship Program grant \#HST-HF2-51453.001 awarded by the Space Telescope Science Institute, which is operated by the Association of Universities for Research in Astronomy, Inc., for NASA, under contract NAS5-26555.
K.Cr. is supported by a UBC Four Year Fellowship (6456).
M.E.D. acknowledges support from the Naval Research Laboratory by NASA under contract S-15633Y.
T.D. and M.T.L. are supported by an NSF Astronomy and Astrophysics Grant (AAG) award number 2009468.
E.C.F. is supported by NASA under award number 80GSFC21M0002.
G.E.F., S.C.S., and S.J.V. are supported by NSF award PHY-2011772.
K.A.G. and S.R.T. acknowledge support from an NSF CAREER award \#2146016.
The Flatiron Institute is supported by the Simons Foundation.
S.H. is supported by the National Science Foundation Graduate Research Fellowship under Grant No. DGE-1745301.
N.La. acknowledges the support from Larry W. Martin and Joyce B. O'Neill Endowed Fellowship in the College of Science at Oregon State University.
Part of this research was carried out at the Jet Propulsion Laboratory, California Institute of Technology, under a contract with the National Aeronautics and Space Administration (80NM0018D0004).
D.R.L. and M.A.Mc. are supported by NSF \#1458952.
M.A.Mc. is supported by NSF \#2009425.
C.M.F.M. was supported in part by the National Science Foundation under Grants No. NSF PHY-1748958 and AST-2106552.
A.Mi. is supported by the Deutsche Forschungsgemeinschaft under Germany's Excellence Strategy - EXC 2121 Quantum Universe - 390833306.
P.N. acknowledges support from the BHI, funded by grants from the John Templeton Foundation and the Gordon and Betty Moore Foundation.
The Dunlap Institute is funded by an endowment established by the David Dunlap family and the University of Toronto.
K.D.O. was supported in part by NSF Grant No. 2207267.
T.T.P. acknowledges support from the Extragalactic Astrophysics Research Group at E\"{o}tv\"{o}s Lor\'{a}nd University, funded by the E\"{o}tv\"{o}s Lor\'{a}nd Research Network (ELKH), which was used during the development of this research.
S.M.R. and I.H.S. are CIFAR Fellows.
Portions of this work performed at NRL were supported by ONR 6.1 basic research funding.
J.D.R. also acknowledges support from start-up funds from Texas Tech University.
J.S. is supported by an NSF Astronomy and Astrophysics Postdoctoral Fellowship under award AST-2202388, and acknowledges previous support by the NSF under award 1847938.
C.U. acknowledges support from BGU (Kreitman fellowship), and the Council for Higher Education and Israel Academy of Sciences and Humanities (Excellence fellowship).
C.A.W. acknowledges support from CIERA, the Adler Planetarium, and the Brinson Foundation through a CIERA-Adler postdoctoral fellowship.
O.Y. is supported by the National Science Foundation Graduate Research Fellowship under Grant No. DGE-2139292.

\bibliography{main}{}
\bibliographystyle{aasjournal}

\end{document}